\documentclass{article}
\usepackage{graphics}

\makeatletter    
\def\eqalign#1{\null\,\vcenter{\openup\jot\m@th \let\\=\crcr 
  \ialign{\strut\hfil$\displaystyle{##}$&$\displaystyle{{}##}$\hfil
      \crcr#1\crcr}}\,}
\makeatother

\def\pmb#1{\setbox0=\hbox{$#1$}%
  \kern-.025em\copy0\kern-\wd0
  \kern.05em\copy0\kern-\wd0
  \kern-.025em\raise.0433em\box0}

\def\bfbeta{\pmb{\beta}} 

\def\bfone{\hbox{\bf 1}}
\def\bfA{\hbox{\bf A}}
\def\bfB{\hbox{\bf B}}
\def\bfe{\hbox{\bf e}}
\def\bfg{\hbox{\bf g}}
\def\Hscr{{\cal H}}
\def\Mscr{{\cal M}}
\def\Tscr{{\cal T}}
\def\Gscr{{\cal G}}

\def\diag{\mathop{\rm diag}\nolimits}

\def\Tr{\mathop{\rm Tr}\nolimits} 
\def\fraction#1#2{{\textstyle\frac#1#2}}
\def\half{\fraction12}

\def\journalfont{\it}         
\def\jou#1{{\journalfont #1\ }}
\def\joudef#1#2{\def #1{\jou{\ignorespaces #2}}}

\joudef{\aaa}{  Astron.\ Astrophys.}
\joudef{\aip}{  Adv.\ Phys.}
\joudef{\am}{   Ann.\ Math.}
\joudef{\ap}{   Ann.\ Phys.\ (N.Y.)}
\joudef{\aop}{   Ann.\ Phys.\ (N.Y.)}
\joudef{\apj}{  Astrophys.\ J.}
\joudef{\cjp}{  Can.\ J.\ Phys.}
\joudef{\cmp}{  Commun.\ Math.\ Phys.}
\joudef{\cqg}{  Class.\ Quantum Grav.}
\joudef{\grg}{  Gen.\ Relativ.\ Grav.}
\joudef{\ijtp}{ Int.\  J.\ Theor.\ Phys.}
\joudef{\jmp}{  J.\ Math.\ Phys.}
\joudef{\jpamg}{ J.\ Phys.\ A: Math.\ Gen.}
\joudef{\mnras}{ Mon.\ Not.\ R.\ Ast.\ Soc.}
\joudef{\nat}{  Nature}
\joudef{\ncim}{ Nuovo Cim.}
\joudef{\nucp}{ Nuc.\ Phys.}
\joudef{\ncb}{  Il Nuovo Cimento ``B}
\joudef{\pl}{   Phys.\ Lett.}
\joudef{\pr}{   Phys.\ Rev.}
\joudef{\prep}{ Phys.\ Rep.}
\joudef{\prl}{  Phys.\ Rev.\ Lett.}
\joudef{\ptp}{  Prog.\ Theor.\ Phys.}
\joudef\rmp{  Rev.\ Mod.\ Phys.}
\joudef\spj{  Sov.\ Phys.\ JETP}

\joudef{\AAA}{  Astron.\ Astrophys.}
\joudef{\AIP}{  Adv.\ Phys.}
\joudef{\AM}{   Ann.\ Math.}
\joudef{\AP}{   Ann.\ Phys.\ (N.Y.)}
\joudef{\AOP}{   Ann.\ Phys.\ (N.Y.)}
\joudef{\APJ}{  Astrophys.\ J.}
\joudef{\CJP}{  Can.\ J.\ Phys.}
\joudef{\CMP}{  Commun.\ Math.\ Phys.}
\joudef{\CQG}{  Class.\ Quantum Grav.}
\joudef{\GRG}{  Gen.\ Relativ.\ Grav.}
\joudef{\IJMP}{ Int.\  J.\ Mod.\ Phys.}
\joudef{\IJTP}{ Int.\  J.\ Theor.\ Phys.}
\joudef{\JKPS}{  J.\ Korean.\ Phys.\ Soc.}
\joudef{\JMP}{  J.\ Math.\ Phys.}
\joudef{\JPAMG}{ J.\ Phys.\ A: Math.\ Gen.}
\joudef{\MNRAS}{ Mon.\ Not.\ R.\ Ast.\ Soc.}
\joudef{\NAT}{  Nature}
\joudef{\NCIM}{ Nuovo Cim.}
\joudef{\NUCP}{ Nuc.\ Phys.}
\joudef{\NCB}{  Il Nuovo Cimento ``B}
\joudef{\PL}{   Phys.\ Lett.}
\joudef{\PR}{   Phys.\ Rev.}
\joudef{\PREP}{ Phys.\ Rep.}
\joudef{\PRL}{  Phys.\ Rev.\ Lett.}
\joudef{\PTP}{  Prog.\ Theor.\ Phys.}
\joudef\RMP{  Rev.\ Mod.\ Phys.}
\joudef\SPJ{  Sov.\ Phys.\ JETP}

\def\vol#1{{\bf #1}}

\def\tabrule{\noalign{\hrule}}

\def\C#1#2#3{C^{#1}{}_{#2#3}}
\def\gl{\hbox{\sl g}}
\def\gm{\hbox{\rm g}}
\def\bfgt{\tilde{\bfg}{}}
\def\bfE{\hbox{\bf E}}
\def\bfR{\hbox{\bf R}}

\def\alphab{\overline{\alpha}}
\def\betab{\overline{\beta}}

\def\lad{(\ln A){} \,\dot{}\,}
\def\lbd{(\ln B){} \,\dot{}\,}
\def\na{N^2A^{-2}}

\def\gee{{}^4 G^{\bot}{}_{\bot}}
\def\nthree{n^{(3)2}}
\def\none{n^{(3)}n^{(1)}}
\def\vvbox#1#2{\vbox{\hbox{\strut#1\hfil} \hbox{\strut#2\hfil}}}
\def\rone{ {}^4 R_A }
\def\rthree{ {}^4 R_B} 

\begin{document}

\setbox43=\vbox{\small\noindent
Reformatted reprint from\\
{\it Origin and Early History of the Universe\/},
 {\it Proc.\ 26th Li\`ege Int.\ Astrophys.\ Colloq. (1986)\/},
edited by J. Demaret  (University of Li\`ege Press, Li\`ege, 1986),
p.~237--254.}

\title{\leavevmode\phantom{H}\\
Higher Dimensional Cosmological Models:\\ the View from Above}

\author{
Robert T. Jantzen\\
\it Department of Mathematical Sciences\\
\it  Villanova University, Villanova, PA 19085 USA \\
\it and\\ 
\it International Center for Relativistic Astrophysics\\
\it Department of Physics, University of Rome, 00185 Rome, Italy}

\date{
}
\maketitle

\vskip -3.25in\box43
\vskip3.25in

\begin{abstract}
A broader perspective is suggested for the study of higher
dimensional cosmological models.
\end{abstract}

\section{Introduction}

In this context one might imagine ``the view from above" to refer to the
vantage point of the extra dimensions of spacetime somewhere ``up there"
above us, as in the fiber bundle diagrams one often sees in discussions
of higher dimensional theories.  However, in using this phrase here I
have in mind something quite different.

No matter how many extra dimensions we consider for spacetime, they all
live in a 2-dimensional world of paper, blackboards, computer screens,
overhead transparencies, $\ldots$  \ In fact an incredible quantity of
2-dimensional space has been filled with discussions of these higher
dimensional models.   It is often easier to keep grinding out the next
small step in this production line, but sometimes one has to step back
and take a more global look at the situation, to rise up above this
2-dimensional detail (in our 3-dimensional world) and think about broader
questions.  How do things fit together?  Are there properties which are
general for the many special cases considered?  What properties of these
special cases will continue to have validity in a more general setting?

In these brief remarks I cannot say many things which deserve being said.
Being somewhat of an outsider in the field, I must also be careful not to
say things which reveal my own ignorance of and lack of familiarity with
much of the current work in higher dimensional models.  Having recently
studied some properties of classical symmetric cosmological models in
higher dimensions,$^1$ this invited talk pushed me to try to make some sense
of some of the details scattered in the literature.

Since time has been short, I cannot explain exactly how everything fits
together, but I hope to convey a general picture which if pursued may put
many of these results into perspective.
In so doing I would suggest that perhaps a little more attention should be
paid to how specific calculations fit into the larger scheme of things.
Certainly it is much more satisfying when many little things fit together
 as particular cases of a single bigger thing.  This idea characterizes
present day theoretical physics and the importance it gives to unifying
symmetries. 

Unfortunately the search for exact solutions of 4-dimensional gravitational
theories has more often than not focused on special cases without putting them
into perspective.  In higher dimensions there is much more room to play games.
I think the situation demands some restraint.  Some effort should be made toward
``looking at the forest rather than the trees."  The game of looking for
exact or even qualitative solutions of gravitational field equations all too
often finds itself at the tree level when an aerial view of the forest is
really what is needed.

\begin{figure}[p]
\resizebox{343pt}{351pt}{\includegraphics{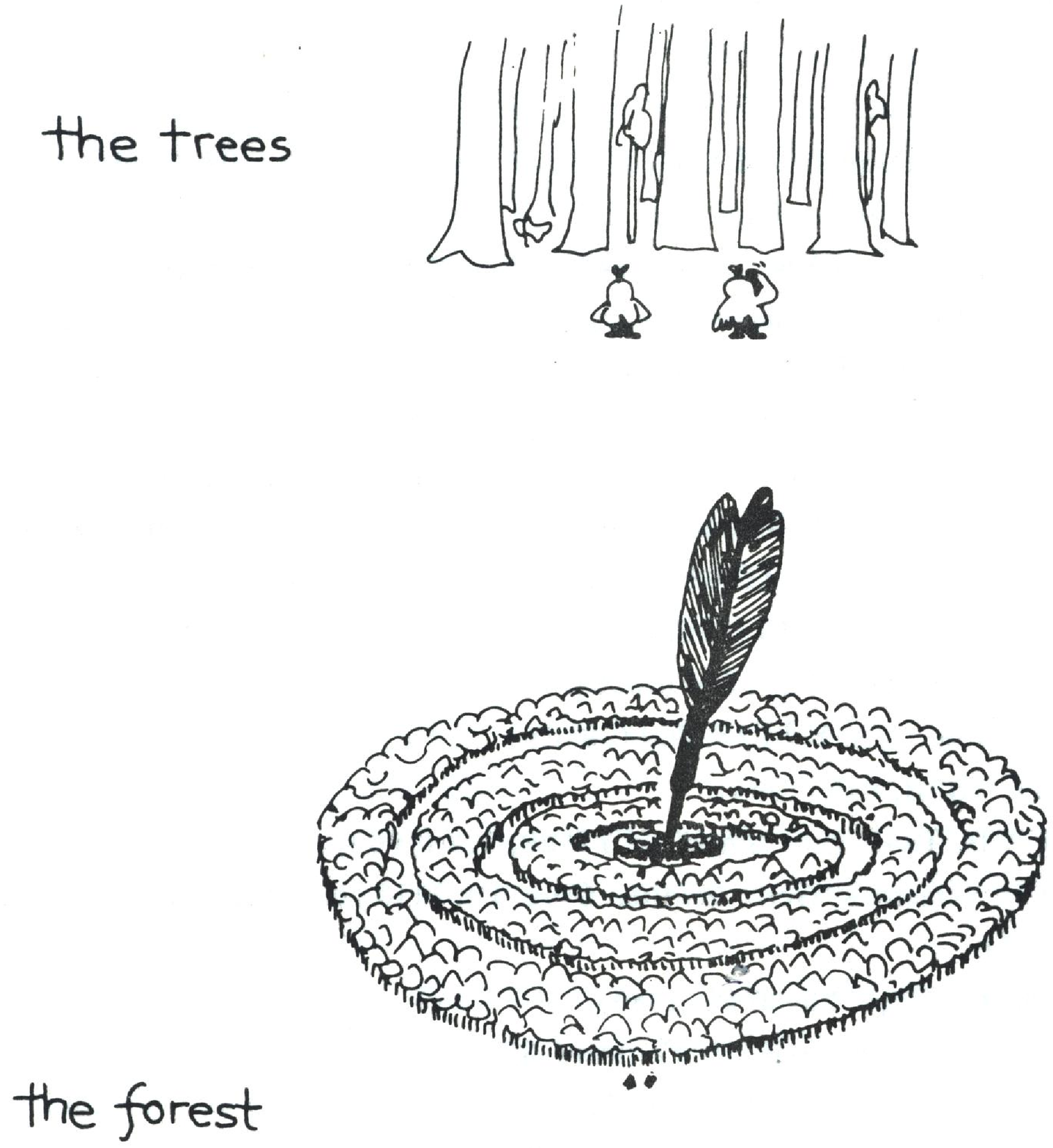}}
\caption{``...not seeing the forest for the trees..." [English language saying]}
\end{figure}

Unification of the fundamental forces has led people to adopt the
Kaluza-Klein idea$^{2-5}$ 
of formulating theories on higher dimensional spacetimes.
At first the extra dimensions were introduced as a mathematical device for
obtaining dimensionally reduced unified theories on 4-dimensional spacetime.
These additional dimensions, tied to symmetries of the spacetime, allowed 
the collapse of the theory to an equivalent 4-dimensional one.  Eventually
ideas progressed to the point where the extra dimensions were taken seriously.
The ``fiber" symmetry of these fiber bundle formulations of 4-dimensional
theories was abandoned and such theories became truly higher dimensional 
ones.$^{6,7}$

This immediately created a problem.  If the fundamental theory of matter is
really higher dimensional in such a nontrivial way, one must explain the
present effective 4-dimensional spacetime and the apparently 4-dimensional
theory which describes nature.  Some mechanism is required to collapse the
additional dimensions to a point where they are not observable.  This could
be accomplished if they are associated with spatially compact dimensions
with very small circumferences, but one needs some way for this 
``compactification" to be a natural consequence of the theory itself.

An interesting possibility for this compactification mechanism is the
dynamical behavior of the gravitational field itself within these theories.
The present 4-dimensional phase of the universe is very anisotropic, given
the necessarily large difference in length scales associated with the 
``ordinary" and ``extra" dimensions.  Chodos and Detweiler$^8$ first
suggested this anisotropy might be a consequence of the evolution of the
gravitational field.  Examining purely gravitational 5-dimensional
theories with flat
homogeneous space sections characterized entirely by ``free" anisotropic
expansion,$^9$ they noted that classical cosmological solutions are
{\it allowed} in which preferential expansion of three spatial
directions occurs.  However, such solutions are not {\it selected}
by the theory.  One needs matter fields to obtain such a selection 
effect.$^{10}$  Such fields may be added in ordinary Kaluza-Klein theories,
while supersymmetry dictates what matter fields exist in locally supersymmetric
theories (supergravity, superstrings).$^{11-19}$

With this in mind classical solutions of the Bose sector of higher dimensional
theories have been studied to throw some light on the nature of the ground state
of such theories.  Such studies fall into two categories, characterized by the
behavior of the gravitational field.  In the first, solutions are considered in 
which the spacetime is the product of two static symmetric component manifolds
representing the ordinary spacetime and the internal spaces, the discussion
being more particle physics in flavor (e.g., reference 13).
In the second, evolving models are
considered, typically from the viewpoint of a relativist somewhat less
concerned with the particle physics motivation for considering such models.

As a classical relativist I wish to address the second kind of investigation
in the spirit of Freund,$^{10}$ Demaret et al$^{20,21}$ and others.$^{22-27}$
Many particular exact or qualitative
cosmological solutions of various higher dimensional theories have been
discussed.  Although only for very special conditions can one obtain closed form
solutions, such solutions may describe a single isolated
dynamical effect of the full equations which when accompanied by other degrees
of freedom no longer allows exact solutions but perhaps exhibits a similar
qualitative behavior, or at least one which can be understood in the context
of the coupled behavior.  Properties exhibited by special solutions are
probably only interesting to the extent that they relate to the general case
in this way.

Certain exact solutions in fact exhibit a ``typical behavior" during certain
phases of the evolution.  Such solutions can be understood in terms of
additional properties related to a qualitative formulation of the field
equations.  From a purely mathematical viewpoint, the existence of closed
form exact solutions often is accompanied by underlying properties of the
differential equations that are rarely understood in first obtaining them and
which help to relate various particular solutions to each other.  Understanding
how exact solutions fit into a bigger picture is perhaps more important than
any particular exact solution found by chance or ``cleverness".

These ideas are also useful in understanding the idea of ``gravitational
chaos" both in the general cosmological context and in the context of the
homogeneous models.$^{28-31,23-26}$  
This question is studied as a purely gravitational
phenomenon relevant to the classical
initial singularity independent of particle theory questions.
Although quantum effects at the Planck length probably
make the question irrelevant, it is an interesting mathematical problem.

Limitations of time and space permit only a hint of a more global point of
view illustrated by a few key examples.  Hopefully such a discussion,
necessarily abbreviated, can still be valuable in establishing the importance
of such a point of view for this branch of cosmology.

\section{ Spatially Homogeneous Models}

The kinds of cosmological solutions considered in higher  dimensional theories
in connection with possible properties of the ``ground state" all share one
feature: they are spatially homogeneous.$^{32}$  
The $d$-dimensional spacetime
contains a 1-parameter family of $(d-1)$-dimensional spacelike hypersurfaces
(let $D=d-1$) which are orbits of an isometry group $G$.  The space sections
are therefore ``homogeneous spaces" diffeomorphic to the coset space $G/H$,
where $H$ is the isotropy group at some point of spacetime, namely the
subgroup of isometries which leave that point fixed.

When $H$ is the identity subgroup or a discrete invariant (i.e., normal)
subgroup, ``space" is isometric to a group manifold (namely the group $G/H$)
with a Riemannian metric invariant under either the left or right translations
of the group into itself, the choice being a matter of convention (left in this
article).  If $H$ is discrete but not invariant, this is still true locally.
All of these cases may be referred to as the simply transitive case, where
$G$ acts as a $D$-dimensional translation group of a $D$-dimensional 
Riemannian manifold.

In the multiply transitive case, $H$ is continuous and $G$ acts as a group of
translations and certain rotations of the space sections, each of which is
isometric to the left coset space $G/H$ with a Riemannian metric invariant
under the natural left action of $G$.  If $H$ is an invariant subgroup, $G/H$
is a group and hence this occurs as a special case of the simply transitive
case for which additional symmetry exists; the case where $H$ is not invariant
will be referred to as nontrivial.

In either case the metric on a given spatial section is completely determined by
its value at a single point and hence the field equations reduce to ordinary 
differential equations in the parameter labeling the spatial slices, easily
interpreted as a time function on the spacetime. (This assumes that the matter
fields either share the symmetry of the metric or at least have enough symmetry
to be compatible with the field equations.)  By choosing the coordinate
lines orthogonal to the spatial sections, the spacetime metric  can be
expressed at least locally in the form$^{33}$
$$ {}^{(d)}\gm =- N(t)^2\, dt\otimes dt + g_{ab}(t)\omega^a \otimes \omega^b
\eqno(2.1)$$
on the manifold $R\times (G/H)$, where $\{\omega^a\}$ are
1-forms on $G/H$ dual to the frame $\{e_a\}_{a=1,\ldots,D}$ characterized by
structure functions $\C abc=\omega^a([e_a,e_b])$ on $G/H$.  These are
constants in the simply transitive case where $\{e_a\}$ is a left invariant
frame (i.e., a basis of the Lie algebra $\gl$ of $G$), but functions on
$G/H$ in the nontrivial multiply transitive case.  In the first case the
matrix $\bfg=(g_{ab})$ is an arbitrary positive definite symmetric matrix,
while in the second case it must satisfy additional linear 
constraints.$^{18,19}$
Similar constraints in the first case may lead to additional spatial symmetries.

The choice of lapse function $N(t)$ determines the parametrization of the
family of homogeneous spatial sections.  The proper time $\tau$, defined by
$d\tau =N(t)dt=\omega^{\bot} $ corresponds to unit lapse.  In expanding
models with an initial singularity, $\tau$ is usually chosen so that
this singularity occurs at $\tau=0$.  Defining $\omega^0=dt$ leads to the
spacetime 1-forms $\omega^{\alpha}$ dual to the frame $\{e_{\alpha}\}_{\alpha
=0,1,\ldots,D}$, where $e_0=\partial/\partial t$.

In practice one is usually interested in spatial geometries which decompose
into an orthogonal product of an ordinary 3-dimensional homogeneous space
and a compact internal homogeneous space
$$  G/H\sim M_3\times(G^{int}/H^{int})\ .
\eqno(2.2)$$
$M_3$ is a Bianchi manifold$^{34}$ in the simply transitive (group manifold)
case and the Kantowski-Sachs manifold$^{35}$ in the nontrivial multiply
transitive case, where $M_3$ is of the form $G/H$ with
$G\sim G_1\times SO(3,R)$, $G_1$ the abelian group
$R$ or $S_1$, and $H\sim SO(2,R)$. (One might refer to $M_3$ as a BKS space.)
A classification of symmetry types in the Bianchi case amounts to a 
classification of 3-dimensional Lie groups/algebras;$^{34,36}$
such a classification is summarized in Table 1.

\begin{table}
\def\tabrule{\noalign{\hrule}}

  \newbox\strutboxfifteen
  \setbox\strutboxfifteen=\hbox{\vrule height 10pt depth 5pt width0pt}
\def\strutfifteen{\relax \ifmmode\copy\strutboxfifteen
\else \unhcopy\strutboxfifteen\fi}

\vbox{\tabskip=0pt \offinterlineskip
\halign{\strutfifteen#& \vrule#\tabskip=.5em plus.5em& 
 #\hfil&\vrule#& #\hfil&\vrule#& #\hfil&\vrule#&
\hfil#\hfil&\vrule#& \hfil#\hfil&\vrule# \tabskip=0pt\cr 
\omit\hfil&\omit\hfil & &\multispan9 \tabskip=0pt \hrulefill\cr 
&\omit\hfil & BIANCHI TYPES & & \multispan3 nonsemisimple \hfil& &
\multispan3 \hfil semisimple & \cr
\omit\hfil&\omit\hfil & &
\multispan9 \tabskip=0pt \hrulefill\cr
&\omit\hfil & & & \multispan5 noncompact \hfil& &
\hfil compact & \cr
\omit\hfil&\omit\hfil & &
\multispan9 \tabskip=0pt \hrulefill\cr
&\omit\hfil & & & abelian & & \multispan5 \hfil nonabelian & \cr\tabrule
& & class $A$ (unimodular)
   & & \hfil I & & II, VI$_0$, VII$_0$ & & VIII & & IX & \cr\tabrule
& & class $B$ (nonumimodular)
   & & & & III, IV, V, VI$_h$, VII$_h$ & & \multispan3 \hfil & \cr\tabrule
}}

\caption{
The isomorphism classes of 3-dimensional Lie groups/algebras.
The compact/noncompact division refers to the simply connected covering
groups of each type, while unimodularity refers to the matrix representation
of the adjoint group of each type.}
\end{table}

\section{Field Equations}

Since we wish to focus on the gravitational properties of higher dimensional
models, we consider the Einstein equations with all other fields lumped into
the energy-momentum tensor.  This assumes that the gravitational part of the
classical field equations can be represented in this way, thus excluding
``higher derivative" theories.  These equations
$$ 0=M_{\alpha \beta}\equiv |{}^{(d)}g|^{1/2} ({}^{(d)}G_{\alpha \beta}-
  T_{\alpha\beta})\ , \qquad |{}^{(d)}g|^{1/2}  = Ng^{1/2} 
\eqno(3.1)$$
(where metric quantities without the leading superscript `${}^{(d)}$' refer
to the spatial metric) naturally split into evolution equations $0=M_{ab}$
which evolve the spatial metric and constraints on the solutions of
those equations
$$ \eqalign{
&0= \Hscr\equiv 2N^{-1}M^{\bot}{}_{\bot} \equiv 2M^0{}_{\bot} \qquad
(\hbox{super-Hamiltonian constraint})\cr
&0= \Hscr_a\equiv 2N^{-1}M^{\bot}{}_a \equiv 2M^0{}_a \qquad
(\hbox{supermomentum constraint})\ .\cr}
\eqno(3.2)$$

The Einstein equations may also be written in ``Ricci form"
$$ 0=P_{\alpha\beta}\equiv {}^{(d)}R_{\alpha\beta} - E_{\alpha\beta}\ ,
\qquad E_{\alpha\beta} \equiv T_{\alpha\beta}-(d-2)^{-1}T^{\gamma}{}_{\gamma}
\,g_{\alpha\beta}\ ,
\eqno(3.3)$$
a form very convenient for static solutions in discussions of spontaneous
compactification.  However, for dynamical solutions a compromise is preferable,
namely the Ricci evolution equations $0=P_{ab}$, supplemented by the Einstein
constraints rather than more equations involving second time derivatives.

To understand the usefulness of this compromise for spatially homogeneous
solutions, it is convenient to introduce the Hamiltonian formulation.
The Einstein evolution equations written in first order form are the
Hamiltonian equations for the spatial metric derived from the Hamiltonian
$H=N\Hscr$ with the lapse treated as an independent variable; here $\pi^{ab}$
is the momentum canonically conjugate to the spatial metric $g_{ab}$
$$\eqalign{
0&=\dot{\pi}{}^{ab} +\partial/\partial g_{ab} (N\Hscr)- NQ^{ab}\cr
 &=\dot{\pi}{}^{ab} +N \partial H/\partial g_{ab} -NQ^{ab} +\partial/\partial
g_{ab} (\ln N) =M^{ab} +H\partial/\partial g_{ab}(\ln N)\ .\cr}
\eqno(3.4)$$
(The term $Q^{ab}$ arises in the case of nonunimodular symmetry groups where
the spatial Einstein tensor force is no longer derivable from the scalar
curvature potential but has a nonpotential component or from similar
problems with the source energy-momentum.$^{37,38}$) If $N$
is treated as independent of $g_{ab}$, the final term vanishes
and one obtains the Einstein evolution equations.
Letting $N$ depend explicitly on $g_{ab}$ leads to evolution equations which
differ by a term proportional to the super-Hamiltonian constraint.
The choice of unit lapse $N=1$ made in proper time time gauge leads to the
Einstein evolution equations, while the choice $N=(const)g^{1/2}$ made in
Taub's time gauge$^{32}$ (Misner supertime$^{39}$ time gauge) leads to a
set of equations which are equivalent to the Ricci evolution equations.

Why is the Taub time gauge useful, i.e., why is it useful to choose the lapse
proportional to the spatial volume factor?  For the answer, one must examine
the differential geometry of the Einstein equations in Hamiltonian form.
The metric configuration space is the space $\Mscr$ of positive definite 
symmetric matrices $\bfg=(g_{ab})$ or a submanifold of this space.
The Hamiltonian
$$ H=N\Hscr = N\Tscr +N U
\eqno(3.5)$$
consists of a kinetic term associated with the metric variables and a
potential energy term composed of a spatial curvature term and a source term
$$\eqalign{
&U = U^{curv} +U^{source}
\ ,\qquad
 \qquad -dU +Q ^{ab}dg_{ab}= -g^{1/2} (G^{ab}- T^{ab}) dg_{ab} \ ,\cr
&U^{curv} =-g^{1/2}R \ ,\qquad \qquad
 \qquad -dU^{curv} +Q^{curv\ ab}dg_{ab}= -g^{1/2} 
G^{ab}dg_{ab}\ , \cr
&U^{source} =-2g^{1/2}T^{\bot}{}_{\bot}
 \ ,\qquad -dU^{source} +Q^{source\ ab}dg_{ab}= g^{1/2} T^{ab} dg_{ab} \ .\cr}
\eqno(3.6)$$
The exterior derivative $d$ is the one on $\Mscr$, while both the source and
the curvature forces may involve nonpotential components. (One may also
simply treat the source force entirely as an external force.)  The
kinetic energy term
$$\eqalign{
N\Tscr &=\fraction{1}{4} N^{-1} \Gscr^{abcd}\dot{g}_{ab}\dot{g}_{cd}
\qquad (\hbox{velocity phase space})\ ,\cr
&= N\Gscr^{-1}_{abcd}\pi^{ab}\pi^{cd} \qquad (\hbox{momentum phase space})\ ,\cr
\pi^{ab}& \equiv \partial/\partial \dot{g}_{ab} (N\Tscr) \qquad\qquad
(\hbox{Legendre transformation})\ ,\cr}
\eqno(3.7)$$
is just the one associated with the DeWitt$^{40}$  metric $\Gscr$ on $\Mscr$,
apart from the rescaling by the lapse function
$$ N^{-1}\Gscr =N^{-1}g^{1/2} (g^{a(c}g^{d)b} - g^{ab}g^{cd})dg_{ab}\otimes
dg_{cd}\ .
\eqno(3.8)$$
Allowing the lapse to have a factor depending explicitly on $g_{ab}$ is
equivalent to conformally rescaling the DeWitt metric by that factor.$^{39}$
In particular the Taub time gauge choice eliminates the spatial volume factor
from the DeWitt metric, thus simplifying the kinetic terms in the
evolution equations as much as possible and also eliminating a
term involving the total kinetic energy which appears in the Einstein form
of those equations
due to the derivative of the spatial volume factor.
This explains the advantage of using the Ricci evolution equations
accompanied by the Taub time gauge.

Neglecting the potential energy, equivalent to restricting one's attention
to the vacuum abelian case where the spatial curvature vanishes, the evolution
equations describe geodesics of the DeWitt metric $\Gscr$ in proper time
time gauge and geodesics of the rescaled DeWitt metric $g^{-1/2}\Gscr$ in
Taub time gauge.  The Hamiltonian constraint in this case requires that the
kinetic energy vanish and therefore be a null vector.  (In fact the null
geodesics are conformally invariant.)

This immediately raises the question of the signature of the DeWitt metric.
It is a Lorentz metric which endows $\Mscr$ with a natural splitting into
space and time which itself plays an important role in the geometry of the
Einstein equations. Let
$$ \bfg =e^{2\alpha}\tilde{\bfg}\ , \qquad g=\det \bfg =e^{2D\alpha}\ ,
\qquad \det \tilde{\bfg}=1
\eqno(3.9)$$
decompose $ \bfg$ into the ``scale invariant" conformal metric $\tilde{\bfg}$
and the conformal scale factor $e^{\alpha}$.  Dynamically in the spatially
homogeneous models $e^{\alpha}$ describes the overall expansion or contraction
of the spatial sections while $\tilde{\bfg}$ describes the anisotropy. The
DeWitt metric in these new variables is explicitly Lorentzian with $\alpha$
playing the role of a natural time variable
$$ g^{-1/2}\Gscr =-D(D-1)d\alpha \otimes d\alpha +
\Tr(\bfgt^{-1}d\bfgt \otimes \bfgt^{-1}d\bfgt) \ .
\eqno(3.10)$$
The second term is Riemannian since at $\bfgt=\bfone$ it leads to the trace
of the square of a symmetric matrix and such an inner
product is positive definite on the space of symmetric matrices.

The diagonal submanifold $\Mscr_D \subset \Mscr$ is in fact a flat Lorentz
spacetime with respect to the rescaled DeWitt metric.  At diagonal points
$\bfgt_D$ one may define $\bfbeta\equiv \half\ln \bfgt_D$ and introduce a basis
$\{\bfe_A\}_{A=1,\ldots,D-1}$ of tracefree diagonal matrices normalized by
the condition $\Tr \bfe_A\bfe_B=D(D-1)\delta_{AB}$, and let $\bfbeta=
\beta^A\bfe_A$, in which case $\{\alpha,\beta^A\}$ are orthonormal
cartesian coordinates with respect to the suitably rescaled DeWitt metric
$$ [D(D-1)g^{1/2}]^{-1}\Gscr|_{\Mscr_D}
 = -d\alpha\otimes d\alpha + \delta_{AB}
d\beta^A \otimes d\beta^B\ .
\eqno(3.11)$$
The vacuum abelian diagonal dynamics in Taub time gauge has as solutions
straight null lines in $\alpha$-$\bfbeta$ space (unit $\alpha$-speed
trajectories in $\bfbeta$ space) with these orthonormal coordinates linear 
in the Taub time $t$.  (It is convenient to choose $N= 2D(D-1)g^{1/2}$ to
obtain the usual factor of one half in the kinetic energy.)
The diagonal components of $\bfg$ are therefore exponential in the Taub time.
Transforming to proper time time gauge leads to power law dependence on the
proper time $\tau$
$$ \bfg_D =e^{\bfA t +\bfB}= \bfg_{D(0)} \diag(\tau^{2p_1},\ldots,
\tau^{2p_D})\ ,
\eqno(3.12)$$
the latter form recognizable as the famous Kasner solution, the Kasner
exponents satisfying the well known identities
$$ \sum\nolimits^D_{a=1} p_a =\sum\nolimits^D_{a=1} p_a{}^2 =1\ .
\eqno(3.13)$$

The Ricci evolution equations in Taub time gauge for diagonal metrics are
explicitly
$$ (\ln \bfg)\, \ddot{}\, =-[2D(D-1)]^2 \cdot 2g(\bfR-\bfE)
\eqno(3.14)$$
where $\bfR= (R^a{}_b)$ and $\bfE= (E^a{}_b)$.  When the righthand side is 
absent, one has free motion with the metric exponential in the Taub time.
This remains true for those degrees of freedom along which the force term
on the righthand side has no component.

As an aside, it is worth mentioning how spatial curvature and nonzero
source energy density change the causal character of the solution curves.
The super-Hamiltonian constraint $\Tscr=-U$ determines the square of the
velocity (or momentum) of the system.  In the case of positive source
energy density, $U$ may be negative and the motion spacelike only for
sufficiently positive spatial curvature.  In a contracting phase this can
lead to a ``bounce" rather than a cosmological singularity, while in an
expanding phase it can lead to a point of maximum expansion and recollapse.
Such ``turnaround phases" are not allowed when the spatial curvature is
negative and the source energy density is positive, where $U$ is
positive and the motion always timelike, with $g$ or equivalently $\alpha$
monotonically increasing (expansion) or decreasing (contraction), respectively
starting or ending at the ``frontier" $g=0$ ($\alpha\to-\infty$), where
the spatial metric is singular.$^{40}$ 
The intersection of a trajectory with the
frontier may or may not represent a physical cosmological singularity.
The variable $\alpha\in(-\infty,\infty)$ is a natural time variable for
$\Mscr$; during a phase of monotonic expansion or contraction, one can choose
it as the time variable $t$ on the underlying spacetime, a gauge introduced
by Misner$^{41}$ ($\alpha$-time time gauge).

\section{Exact Solutions}

A useful example to illustrate some of these ideas and others are the Taub
solutions$^{32}$ in $d=4$ spacetime dimensions. This class of metrics is
defined by
$$\eqalign{& \C123=\C231=n^{(1)}\ , \qquad  \C312=n^{(3)}\ , \qquad
\bfg_D=e^{2\alpha}e^{2(\beta^+\bfe_+ + \beta^-\bfe_-)}\ ,\cr &
\bfe_+=\diag(1,1,-2)\ , \qquad 
\bfe_- =\sqrt3 \diag(1,-1,0)\ ,\qquad \beta^-= \delta(n^{(1)}) \beta^-_{0}\ ,
\cr}
\eqno(4.1)$$
and their dynamics in Taub time gauge
is described by the Hamiltonian (expressed in velocity
phase space)
$$ H=\half (-\dot{\alpha}{}^2 +\dot{\beta}{}^{+\,2} +\dot{\beta}{}^{-\,2})
+6\nthree e^{4(\alpha-2\beta^+)} -24\none e^{-2(\beta^+-2\alpha)}\ .
\eqno(4.2)$$
The Lorentz transformation$^{39}$
$$ (\alphab,\betab{}^+)=3^{-1/2}(2\alpha-\beta^+,-\alpha+2\beta^+)
\eqno(4.3)$$
leads to a completely decoupled Hamiltonian
$$\eqalign{ H&= \half (-\dot{\alphab}{}^2 +\dot{\betab}{}^{+\,2} 
  +\dot{\beta}{}^{-\,2})
 +6\nthree e^{-4\sqrt3\betab{}^+} -24\none e^{2\sqrt3 \alphab}\cr
 & = -H_{\alphab} +H_{\betab{}^+} +H_{\beta^-} =0\cr}
\eqno(4.4)$$
consisting of three 1-dimensional scattering problems with exponential
potentials and constant energies, restricted only by the constraint that
the appropriately signed sum of the individual energies vanish.  The
solution of an exponential scattering problem leads to hyperbolic or
trigonometric functions
$$\eqalign{& H_x= \half(\dot{x}{}^2 +\mu e^{\nu x}) =E_x
\quad\rightarrow\quad t=\int dx (2E_x-\mu e^{\nu x})^{-1/2}\ ,\cr
&\qquad \hbox{or defining $\gamma=(2E_x)^{1/2}$}\ ,\cr
&e^{-{\half}\nu x}=\cases{ \mu^{1/2}
\gamma^{-1} \cosh(\half\gamma\nu t) & $\mu>0$
\ ,\qquad $E_x>0$ \cr
|\mu|^{1/2}
\gamma^{-1} \sinh(\half\gamma|\nu| t) 
& $\mu<0$\ ,\qquad $E_x\in \Re$\ .\cr} \cr}
\eqno(4.5)$$

$(\overline{\alpha},\overline{\beta}{}^+,\beta^-)$ are inertial coordinates
of the rest frame of the $n^{(3)\,2}$ potential, which moves with speed
$d\beta^+/d\alpha=\half$ in $\bfbeta$ space.  For the Bianchi type IX case,
characterized by $n^{(3)}n^{(1)}>0$, the hyperbolic cosine solution is relevant,
interpolating between the asymptotic free positive and negative exponential
solutions (for the metric components) at $t=\pm\infty$; the unit velocities
in $\bfbeta$ space of the asymptotic solutions are related by a simple
reflection in the rest frame of this potential.  Letting $n^{(1)}\to0$
contracts the group to Bianchi type II, eliminating the ``tachyonic"
$n^{(3)}n^{(1)}$ potential (it moves with speed $d\beta^+/d\alpha=2$ in
$\bfbeta$ space) and allows free motion parallel to the $n^{(3)\,2}$
potential (the $\overline{\alpha}$ and $\overline{\beta}{}^-$ directions).
For the Bianchi type VIII case, characterized by $n^{(3)}n^{(1)}<0$, only
the positive energy solutions are relevant for the $\overline{\alpha}$ motion
(the potential is negative) due to the super-Hamiltonian constraint.


The same problem may be viewed directly from the Einstein equations.$^{42}$
Defining $N^2=A^mB^n$, $\bfg_D=\diag(A,A,B)$, ${}^4 R^1{}_1=
{}^4 R^2{}_2={}^4 R_A$, and ${}^4 R^3{}_3={}^4 R_B$
for the locally
rotationally symmetric case, one has
$$\eqalign{
2N^2\, (\gee) &= -\lad \lbd -\half \lad{}^2 +\na (\half\nthree B- 2\none A)
\ ,\cr
2N^2 ({}^4 R_A) &= (\ln A)\,\ddot{}\, -\lad (\ln NA^{-1} B^{-1/2})\, 
\dot{}\,  +\na (-\nthree B +2\none A)\ ,\cr
2N^2 ({}^4 R_B) &= (\ln B)\,\ddot{}\, -\lbd (\ln NA^{-1} B^{-1/2})\, 
\dot{}\,  +\na (\nthree B)\ .\cr}
\eqno(4.6)$$
Different choices of $(m,n)$ lead to decoupling of different products of
$A$ and $B$.  Taub's$^{32}$ choice $(2,1)$ decouples $(AB)^{-1}$ and $B$, 
Brill's$^{43}$ choice $(2,-1)$ decouples $A^{-1/2}$ and $B$, while
Misner's$^{44}$  choice $(0,-1)$
decouples $AB$ and $A$, the special exponents yielding
the simplest form for the solutions.
However, the decoupling
of the evolution equations requires adding appropriate multiples of
the super-Hamiltonian constraint to the Einstein evolution equations.
Table 2 summarizes the decoupling possibilities for the evolution
equations for the semisimple case.

\begin{table}
\vbox{\tabskip=0pt \offinterlineskip
\halign{\strut#& \vrule#\tabskip=.5em plus.5em&
#\hfil& \vrule#& #\hfil& \vrule#& #\hfil& \vrule#&
#\hfil& \vrule#& #\hfil& \vrule#&  #\hfil& \vrule#\tabskip=0pt\cr \tabrule
& & \vvbox{evolution}{equation (set =0)} & &
 \vvbox{spatial}{curvature} & &
 \vvbox{2nd time}{derivative} & &
 \vvbox{quadratic}{decoupling} & &
 \vvbox{quadratic}{elimination} & &
 \vvbox{linear}{choice} & \cr \tabrule
& & $\rone +\gee$ & & $\nthree$ & & A & & $n=-1$ & & $m=1$ & & $(1,-1)$ &\cr 
\tabrule
& & $\rthree$ & & $\nthree$ & & B & & $m=2$ & & $n=1$ & & $(2,1)$ &\cr \tabrule
& & $\rthree-2(\gee)$ & & $\none$ & & B & &  $m=2$ & & $n=1$ & & $(2,1)$ &\cr 
\tabrule
& & $\rthree+\rone$ & & $\none$ & & AB & & $m=n+1$ & & $n=1$ & & $(2,1)$ &\cr 
\tabrule
& \multispan{13}\hfil\cr
}}
\caption{}
\end{table}

\begin{table}

\vbox{\tabskip=0pt\offinterlineskip
\halign{\strut#& \vrule#\tabskip=.5em plus .5em&
 #\hfil& \vrule#& #\hfil& \vrule#& #\hfil& \vrule#\tabskip=0pt\cr \tabrule
& & independent variables & & lapse choice & & constant of motion &\cr \tabrule
& & $(B,AB)$ & & $(2,1)$: Taub & & $A^2B(\gee)$ &\cr \tabrule
& & $(A,B)$ & & $(2,-1)$: Brill & & $A(\gee)$ &\cr \tabrule
& & $(A,AB)$ & & $(0,-1)$: Misner & & $A^2(\gee)$ &\cr \tabrule
& \multispan7\cr
}}
\caption{}
\end{table}

The first column of Table 2 lists those linear combinations of the three
independent spacetime curvatures $\rone$, $\rthree$, and $\gee$  which contain
only one of the two independent spatial curvatures $ R^1{}_1 + R^3{}_3$
($\sim \none$) and $R^3{}_3$ ($\sim \nthree$).  The second column
indicates which spatial curvature appears while the third column indicates
the variable whose natural log appears in the second derivative term.
The fourth column gives the condition that the quadratic derivative term
only involve this variable and the fifth column gives the condition that the
quadratic term be absent.  The last column gives the values of $(m,n)$ for
satisfying both conditions.

Table 3 describes the choice of variables for which both evolution equations
decouple in the semisimple case; the form of the constant of the motion 
may be derived from the evolution equations.
In the nonsemisimple case of Bianchi type II
($n^{(1)}=0,n^{(3)}\ne0$), the vanishing of the $\none$ curvature term leads 
to the
additional lapse choice $(1,-1)$ of Bradley and Sviestens$^{45,46}$ for which
both evolution equations decouple.
When $n^{(1)}=n^{(3)}=0$, one obtains the abelian case of Bianchi type I and the
solutions are equivalent to the Kasner solution expressed with different
time functions.

Why give so much attention to a vacuum solution of $d=4$ general relativity
in an article devoted to higher dimensional cosmological models?  The
answer is that the Taub family of solutions are not content to remain in
four spacetime dimensions but keep reappearing in higher dimensional
models.

For solutions of the Ricci evolution equations, the Taub time gauge
Hamiltonian $H\sim g(\gee)$ is a constant function of the parameters of those
solutions.  For negative values of $H$, one must add something positive to
$H$ to obtain zero.  This corresponds to a source with $E_{ab}=0 \ 
(=E^{\bot}{}_a)$ but $gT^{\bot}{}_{\bot}= H_{source}>0$.  Such a source is
a homogeneous stiff perfect fluid moving orthogonally to the space sections,
which in turn is equivalent to a homogeneous massless scalar field.  The
latter is equivalent by conformal transformations to a solution of the
Brans-Dicke field equations.$^{9}$  In other words, by ``varying the
parameters"$^{42}$
which occur in the solution of the Ricci evolution equations away
from the vacuum super-Hamiltonian constraint values, one obtains new
solutions which may be interpreted as containing either or both of these
equivalent sources or as a solution of the Brans-Dicke theory.  (Such a
variation of parameters applied to other choices of evolution equations
leads to the introduction of a locally rotationally symmetric electromagnetic
field source and the Brill generalization of the Taub solutions,$^{43}$
to which stiff perfect fluids or scalar fields may be added,$^{42}$
or a cosmological constant.$^{47}$)

These are still $d=4$ solutions.  However, the massless scalar field source
or Brans-Dicke theory is equivalent to a $d=5$ Kaluza-Klein theory with an
additional flat dimension.$^{9}$  More scalar fields lead to higher
dimensional Kaluza-Klein models, always with flat extra dimensions.

In these examples, essentially only gravitational degrees of freedom have
been considered.  As an example of a model with nongravitational degrees
of freedom, consider $d=11$ supergravity with the Freund-Rubin ansatz.$^{10,11}$
Although already out of fashion, it is a cute example of another appearance
of the Taub solutions.  Spatially homogeneous solutions of the bosonic sector
of the theory have been considered by Freund,$^{10}$ Demaret et al$^{20,21}$
and Lorentz-Petzold.$^{22}$  The spacetime is assumed to be of the form
$M_4\times M_7$, where $M_4\sim R\times M_3$ and the spatial 
metric is decomposable,
i.e., equivalent to independent metrics on the factor manifolds
$M_3$ and $M_7$ which are orthogonal in the product.  The spatial metric matrix
$\bfg$ is then in block diagonal form with matrix blocks $\bfg_3$ and $\bfg_7$,
each of which may be decomposed into scale invariant and scale factor parts
$$ \bfg_3 \leftrightarrow (g_3, \tilde{\bfg}_3)\ , \qquad
 \bfg_7 \leftrightarrow (g_7, \tilde{\bfg}_7)\ .
\eqno(4.7)$$
\def\ab{\overline{a}} 
\def\bb{\overline{b}} 
\def\cb{\overline{c}} 
\def\ah{\overline{a}} 
\def\bh{\overline{b}} 
\def\ch{\overline{c}} 
The Freund-Rubin ansatz which solves the field equations for the 3-form $A$
of the theory assumes that the 4-form $F=dA$  has as its only nonvanishing
components (let $\ab,\bb=1,2,3$)
$$ F^{0\ab\bb\cb} = f_D \, g^{-1/2}_7\, {}^{(4)}\eta^{0\ab\bb\cb}\ ,
 \qquad {}^{(4)}\eta^{0\ab\bb\cb} =
 -(Ng_3^{1/2})^{-1/2} \epsilon^{0\ab\bb\cb}\ ,
\quad \epsilon^{0123}=1\ ,
\eqno(4.8)$$
so that
$$ F_{0\ab\bb\cb} F^{0\ab\bb\cb} =-4! f_D{}^2 g_7{}^{-1}= \fraction{1}{4}
 F_{\alpha\beta \gamma\delta} F^{\alpha\beta \gamma\delta}\ ,
\eqno(4.9)$$
where $f_D$ is a constant (as in Demaret et al$^{20}$).

This leaves an anisotropic source for the Einstein equations. The 
energy-momentum tensor or its Ricci equivalent form
$$ \eqalign{ & T_{\alpha\beta} =\fraction{1}{48}( 8 F^2{}_{\alpha\beta} -
g_{\alpha\beta} F^2)\ , \qquad 
E_{\alpha\beta} =\fraction{1}{6} (F^2{}_{\alpha\beta} -
\fraction{1}{12} g_{\alpha\beta} F^2)\ ,\cr
& F^2{}_{\alpha\beta}\equiv 
 F_{\alpha \gamma\delta\epsilon} F_{\beta}{}^{\gamma\delta\epsilon}\ ,\qquad
F^2\equiv  F_{\alpha\beta \gamma\delta} F^{\alpha\beta \gamma\delta}\ ,\cr}
\eqno(4.10)$$
reduces to
$$ T^0{}_0= -2f_D{}^2g_7{}^{-1} \ , \qquad T^{\ab}{}_{\bb} =-2f_D{}^2g_7{}^{-1}
\delta^{\ab}{}_{\bb}\ ,\qquad T^{\ah}{}_{\bh}= 2f_D{}^2g_7{}^{-1}
\delta^{\ah}{}_{\bh}\ ,
\eqno(4.11)$$
while the source super-Hamiltonian is
$$ U^{source} = 4 f_D{}^2 (g_3/g_7)^{1/2}\ ,\qquad -dU^{source}=g^{1/2}
T^{ab}dg_{ab}\ .
\eqno(4.12)$$
Note that by replacing $F$ by its expression in terms of the metric leads to an
effective potential which does not necessarily yield the correct equivalent
energy-momentum upon variation.  For the Freund-Rubin source no problem arises,
but for a Freund-Rubin-Englert$^{12}$ source, one must introduce a nonpotential
force to compensate.

In Taub time gauge the source potential is 
$$ N U^{source}= (2\cdot10\cdot9) g^{1/2} U^{source} = (2\cdot10\cdot9) 
\cdot 4 f_D{}^2 g_3\ .
\eqno(4.13)$$
Thus only $g_3$ is affected by the source, the variables orthogonal to
$g_3$ continuing to satisfy the vacuum equations.  To discuss the evolution
of the metric one must make assumptions about the homogeneity groups of
$M_3$ and $M_7$.  If the spatial curvature on $M_3$ is such that
the equation for $g_3$ is linear, i.e., the Taub time gauge potential
is linear in $g_3$, then adding the source only changes the coefficient,
i.e., ``varies the parameters" in the evolution equations, which have the
same solutions but a different dependence on the parameters.

In the simplest case $M_3$ and $M_7$ are both flat and the vacuum dynamics
is just ``free motion." With the Freund-Rubin source, the variables
orthogonal to $g_3$ remain free and the single degree on freedom $g_3$
decouples from them leading to the 1-dimensional exponential scattering
problem (in $\ln g_3$); the solution is the Bianchi type II Taub solution
involving one nontrivial curvature term, apart from a variation of parameters.
The variables $g_3$ and $g_7$ are not orthogonal, however.  One must
introduce
$$\eqalign{&\bfg=e^{2(\beta^3\bfe_3 +\beta^{3\bot}\bfe_{3\bot})}
\pmatrix{\tilde{\bfg}_3 & 0\cr 0 & \tilde{\bfg}_7\cr}\ , \qquad
\matrix{ \bfe_3= \diag(1,1,1,-\half,\ldots,-\half) \hfill \cr
 \bfe_{3\bot}= \diag(0,0,0,1,\ldots,1)\hfill\cr} \ , \cr
& g_3= e^{3\beta^3}\ , \qquad g_7= e^{7(-{\half}\beta^3 +\beta^{3\bot})}\ .\cr}
\eqno(4.14)$$
The $\beta^3$ degree of freedom decouples from $\beta^{3\bot}$ which together
with the remaining
anisotropy variables remains free.  The picture in the $\beta^3$-$\beta^{3\bot}$
plane is exactly the Bianchi type II Taub solution with a variation of 
parameters.  In this plane, $\ln g_7$ has slope $\half$; one sees that free
motion in $\beta^{3\bot}$ (parallel to the vertical potential contours) leads to
monotonic expansion or contraction in $g_7$.  The initial asymptotic free
state scatters off the static potential into a final free state with a 
velocity reflected from the potential contour direction, leading to an
expansion from zero to a maximum and then recollapse in $g_3$.
Although this model$^{20}$ is not realistic, it does exhibit phases
of preferential expansion or contraction of the ordinary and extra dimensions.
(The solutions of the Ricci evolution equations for $\bfg_3$ in Taub time gauge
are in fact the same as the $d=4$ spatially flat case with a positive 
cosmological constant, but the proper time is a different function of the Taub
time in the two cases due to the additional dimensions.)

If instead one gives $M_3$ isotropic spatial curvature by letting its
metric be the diagonal Bianchi type V metric, some special solutions may
be obtained from the $d=4$ vacuum solution (found by Joseph$^{48}$) by a
variation of parameters.  This is easily seen by considering this solution
in Taub time gauge $N^2\sim g_3$ instead of the usual time gauge
$N^2=g_{33}$ in which that solution is usually presented
(discussed for spatially homogeneous models by Siklos$^{49}$).
The free motion in the latter time gauge is represented by hyperbolic tangents
instead of exponentials as occur in Taub time gauge, while the single
curvature driven variable is described by hyperbolic sines in both gauges for
this case.  In Taub time gauge one sees that the Joseph solution is again
the Taub solution with one curvature term (Bianchi type II) in disguised
form.

For a diagonal Bianchi type V metric (with $\C \ab\bb\cb = 
a\delta^{3\ab}_{\bb\cb}$)
one has
$$ \bfg_3= e^{2\beta^0} e^{2(\beta^+ \bfe_+ +\beta^-\bfe_-)}\ ,\qquad
 \bfe_+=\diag(1,1,-2)\ , \quad \bfe_-=\sqrt3\diag(1,-1,0)\ ,
\eqno(4.15)$$
but the supermomentum constraint suppresses the $\beta^+$ degree of
freedom (one may set $\beta^+=0$), 
resulting in a Hamiltonian system for the remaining two variables
with $g_{33}=g_3{}^{1/3}$.
In Taub time gauge the curvature potential is
$$ N U^{curv} = (2\cdot10\cdot 9) g_7 g_3 g^{33} a^2\ , =
(2\cdot10\cdot 9) g_7 g_3{}^{2/3} a^2\ .
\eqno(4.16)$$
If one imposes that $g_7$ be proportional to $g_3{}^{1/3}$, this potential
becomes proportional to $g_3$ and simply adds on to the Freund-Rubin
potential. The only difference from the vacuum case is that the coefficient
of $g_3$ in this potential is different; the same general solution is
obtained with a variation of parameters, i.e., the Taub solution occurs again.
Apparently from the work of Demaret et al$^{20}$ and Lorentz-Petzold$^{22}$
this additional condition is compatible with the dynamics.
If one could consistently impose the constraint that this potential depend on
$\beta^{3\bot}$ only, then one would obtain a system equivalent to the 
semisimple
Taub solution which describes the evolution in the presence of two exponential
potentials depending on orthogonal variables.  In  general the two potentials
are not orthogonal and the evolution equations do not decouple in Taub time
gauge; perhaps a direct analysis of the decoupling possibilities with power
law lapses would lead to the solution.

The Joseph
solution is itself a special case of the Ellis-MacCallum diagonal type
VI$_h$ solution,$^{50}$ obtained by Lie algebra contraction of that
solution. The same
variation of parameters extends to this case and the whole family of
$d=4$ Taub solutions, as well as the type VI$_0$ analog of the locally
rotationally symmetric Taub metrics and the Kantowski-Sachs geometry, whose
vacuum solution is related by analytic continuation to the Bianchi
type III (= VI$_{-1}$) value of the Ellis-MacCallum solution.  These are
discussed by Lorentz.$^{22}$ All of these are disguised versions of the
Taub family of solutions which represents the dynamics of one or two
nontrivial gravitational degrees of freedom, independent of the dimension.
This variation of parameter idea is not special to $d=11$ supergravity
in the higher dimensional context, but as already mentioned above, has
implications for other theories as well.  

 
Consider the general case of a decomposable spatial metric from the point of 
view of the Ricci evolution equations
$$ \bfg =\pmatrix{ \bfg_1& 0\cr 0&\bfg_2\cr}\ , \qquad
\bfR-\bfE =\pmatrix{\bfR_1-\bfE_1&0\cr 0&\bfR_2-\bfE_2}\ .
\eqno(4.17)$$
Here $\bfR_1$ and $\bfR_2$ are the Ricci tensor mixed component matrices of 
the individual homogeneous metrics on the factor manifolds $M_1$ and $M_2$; 
the source energy-momentum must also be in block diagonal form for consistency 
with the Einstein equations.  The Ricci evolution equations are
$$ 0= x^{-1}(x \bfg^{-1}\dot{\bfg})\,\dot{} +2N^2(\bfR-\bfE)\ , \qquad 
x\equiv N^{-1}g^{1/2}\ .
\eqno(4.18)$$
and in Taub time gauge $x\sim const$
$$ 0= (\bfg^{-1}\dot{\bfg})\,\dot{} + (2\cdot10\cdot9)^2\cdot
2g_1g_2(\bfR-\bfE)\ .
\eqno(4.19)$$
In the diagonal case $\bfg=\bfg_D$, the time derivative term is simply $(\ln 
\bfg_D)\,\ddot{} $.
 
The only coupling between $\bfg_1$ and $\bfg_2$ in this time gauge occurs 
through the factor $g_1g_2$ in the driving term.  When $\bfR_2-\bfE_2=0$, for 
example, $\bfg_2$ has free dynamics and one can rescale $\bfg_1$ by a power
of $g_2$ to decouple the remaining evolution equations from $g_2$
$$ \eqalign{ & \bfg_1=g_2{}^{\zeta}\overline{\bfg}_1\ , \qquad
\zeta=(D_1-1)^{-1}\ ,\quad D_1=\dim M_1\ ,\cr
 & 0= (\overline{\bfg}{}_1^{-1}\dot{\overline{\bfg}}_1)\,\dot{} + 
(2\cdot10\cdot9)^2\cdot 2\overline{g}_1(\overline{\bfR}_1
-\overline{\bfE}_1)\ .\cr}
\eqno(4.20)$$
This redefinition of variables leads to the Taub time Ricci evolution equations 
for the spacetime $R\times M_1$ with spatial metric matrix 
$\overline{\bfg}_1$.  The conserved energy associated with those equations, 
however, must be nonzero due to 
the contribution of the free motion in the extra dimensions to the higher 
dimensional energy.
This was noted by Barrow and Stein-Schabes for a $d=5$ Kaluza-Klein
generalization of the Bianchi type IX models, for which generalized
Taub solutions were mentioned.$^{23}$
 
When the spatial metric is not decomposable, but still has the block diagonal 
form (4.17), a number of decoupling possibilities occur depending on the form 
of the driving term.  Suppose that $\bfg_2$ is the 1-dimensional matrix 
$(g_{DD})$ so $g_2=g_{DD}$.  Then the gauge $N=g_{DD}{}^{1/2}$ leads to 
$x=g_1{}^{1/2}$; if $N^2(\bfR-\bfE)$ is independent of $g_{DD}$, then $\bfg_1$ 
immediately decouples from $g_{DD}$, while $\ln g_{DD}$ satisfies a linear 
equation with a $\bfg_1$ dependent source.  When the trace of $N^2(\bfR_1
-\bfE_1)$ depends only on $x$, taking the trace of the first block of the 
Ricci evolution equations leads to a decoupled equation for $x$, the kinetic 
term reducing to $2x^{-1}\ddot{x}$.  If $N^2 \Tr(\bfR_1 -\bfE_1)$ is a 
constant, say $-\chi^2$, this equation is in fact $\ddot{x}-\chi^2 x=0$,
leading to exponential or hyperbolic solutions when $\chi^2>0$, trigonometric
solutions when $\chi^2<0$ and linear solutions when $\chi=0$. (This is again 
the 1-dimensional scattering problem, this time in the variable $\ln x$.)
This situation occurs, for example, if the spatial section is the semidirect 
product of a $(D-1)$-dimensional abelian subgroup and a 1-dimensional group of
automorphisms of that subgroup, the 1-dimensional group being associated with 
the $D$\/th direction.  The block diagonal (``symmetric case") $D=3$
nonsemisimple models are of this type, and this feature allows a uniform 
treatment in the time gauge $N=g_{33}{}^{1/2}$
of most of the known vacuum solutions of this class,
which are related by variation of parameters to scalar field (and hence higher
dimensional Kaluza-Klein), stiff
perfect fluid (including tilt along the distinguished 
direction) and spatially self-similar generalizations.$^{42}$  Some solutions 
of the evolution equations, not allowed by the original super-Hamiltonian 
constraint, may satisfy that constraint after a variation of parameters, 
leading to solutions with no analog in the original case.  Such solutions
occur when a certain parameter changes sign.
 
For $x$ to decouple, it is sufficient that $N^2\Tr(\bfR_1-\bfE_1)$ depend only
on $x$. This occurs if this term arises from a curvature term associated with
an isotropic coset space factor manifold to which the $D$\/th direction 
belongs. Friedmann-Robertson-Walker or Kantowski-Sachs factors or even
semisimple Taub factors lead to this behavior.  In the $D=3$ semisimple Taub 
case, the group manifold $S_3$, as an $S_1$ fiber bundle over $S_2$,$^{51}$
is locally a product, hence choosing $N=A^{1/2}$ (i.e., $(m,n)=(1,0)$),
where $A$ is associated with the isotropic $S_2$ base of the fiber bundle,
leads to a decoupled equation for $x=(AB)^{1/2}$. 
One may also rescale the lapse by
a power of $x$ without loosing this decoupling (i.e., $m=n+1$).
For example, in the Taub
case, the lapse $N=x^{-2}A=B^{-1}$ leads to the $(0,-1)$ decoupling of $x$,
yielding quadratic solutions.
 
These ideas were used by 
Lorentz-Petzold (without being fully understood) in his discussion of some
supergravity solutions$^{22}$ and Brans-Dicke solutions.$^{52}$   The latter 
are equivalent to a $d=5$ Kaluza-Klein model in the vacuum case, and his
choice of decoupling variable ``$g$" involving the scalar field is exactly
the variable $x$ in the higher dimensional formulation.  Adding perfect fluids
with certain equations of state leads to a variation of
parameters in the evolution equations without breaking this
decoupling.  All exact solutions can be understood in terms of the variation
of parameters idea.
 
Another situation arises when the lapse depends only on $g$, say 
$N=g^{(1 -\zeta^2)/2}$ so $x=g^{\zeta^2 /2}$.  The trace of the Ricci evolution
equations then gives a decoupled equation for $x$ as long as 
$N^2\Tr(\bfR-\bfE)$ is a constant, say $-\chi^2$; the equation is then
$\ddot{x}-\zeta^2\chi^2 x=0$, again having the same solutions as discussed
above depending on the signs of $\zeta^2$ and $\chi^2$.  For example, in the
$d=4$ case, a Bianchi type I perfect fluid with equation of state 
$p=(\gamma-1)\rho$ has $\bfE\sim g^{-\gamma/2}$ and hence $N\sim g^{\gamma/4}$
makes $N^2\bfE$ constant; a cosmological constant alone has $\bfE$ 
constant, so in this latter case $x$ has
the same solutions as above but in proper time gauge.  On the
other hand for the Bianchi type I fluid case, one can also choose 
$N=g^{(\gamma-1)/2}$ so that $x\sim g^{1-\gamma/2}$ and $N^2\bfE\sim x^{-1}$,
which leads to quadratic solutions for $x$; this was an intermediate step in 
the work of Jacobs.$^{53}$
 
All these exact solutions are characterized by the fact that the source
energy-momentum (including a possible cosmological constant) may be 
represented in terms of the metric and constants of the motion, leading to
an entirely geometric system.  This occurs only when the source has 
either discrete or continuous additional symmetry.$^{37}$
In this class of models at most two ``nontrivial" gravitational modes can
be excited, as in the semisimple Taub case, the existence of exact solutions
depending at least on a partial decoupling of the modes.

\section{Concluding Remarks on Exact Solutions}
 
At this point the overview of existing special exact solutions has shown us 
that they all depend crucially on decoupling ideas that involve the DeWitt 
geometry, the symmetry of the potentials or driving forces, the choice of time 
variable and the choice of evolution equations.  The variation of parameters 
idea and its relation to the constraints turns out to be a very important one,
sometimes involving the parameters of the solutions of the equations and 
sometimes the parameters of the equations themselves. Although the 
super-Hamiltonian constraint has been emphasized because of the models 
discussed here, the supermomentum constraint plays a similar role when models 
are considered which excite the supermomentum.$^{42}$ All of these ideas seem 
never to be systematically applied.  Most special solutions are found almost
by accident and never later understood in a broader context.  Studying the
appropriate sections of the {\it Exact Solutions} book$^{54}$ with these ideas
in mind, one begins to see some order in the seemingly unrelated list of exact
solutions in ordinary spacetime.  In higher dimensions the situation is 
similar.
 
But are particular exact cosmological solutions really significant?  What do 
we gain by finding new very special solutions of some particular set of field
equations?  This is a timely question, given that the exact solutions 
industry often seems to be only weakly coupled to reality.  One might imagine
that special solutions might be interesting if they exhibit some ``stable"
property of the field equations in some sense or if the solutions exhibit some
``typical" behavior during certain phases of the evolution of the universe, at
least within the restricted symmetry class considered if not in a larger 
context.
 
For the spatially homogeneous case, ordinary differential equations permit a
sophisticated analysis of the qualitative behavior of their 
solutions.$^{55-58}$  In an
appropriate formulation of the field equations, certain special exact 
solutions play the role of ``singular points" and ``separatrixes" between
singular points on an appropriate phase space.  In the context of the 
parameter space of all symmetry types of a given dimension, exact solutions of
a lesser symmetry type often characterize phases of evolution of a higher
symmetry type.$^{38}$ Singular points are associated with ``exact power law" 
solutions$^{59,60}$ of the field equations, which in turn are geometrically
characterized by self-similar evolution, i.e., the existence of additional
spacetime symmetry in the form of a homothetic symmetry which shifts the
homogeneous hypersurfaces along the time direction. This reduces the field 
equations to algebraic equations in certain parameters which determine the
metric completely.  The simplest of these is the Kasner solution.  The Taub
solutions and their generalizations act as separatrixes in this picture.  
These kinds of solutions are important in understanding typical behavior at
very early times toward the big bang and at very late times coming out of the
big bang.
 
Perhaps some of the energy devoted to finding particular exact solutions (a 
tree level activity) should be shifted to understanding some of the broader
questions that arise in this area of research.  The average level of ``new
exact solution" papers seems to have remained close to the level of two 
decades ago when this game first drew widespread participation.  As we have 
seen, those exact solutions which exist are very special and have very nice 
mathematical properties which relate them to one another due to the richness 
of the Einstein equations.  If these simple ideas are not commonly understood,
how can one hope to have reasonable progress on more complicated questions?
 
\section{Chaos?}
 
One deeper mathematical question which seems to fascinate many people is the
extent to which the ``oscillatory approach to the initial singularity" found
in the $d=4$ spatially homogeneous vacuum semisimple case and in the 
(nonstiff) perfect fluid case with anisotropic spatial curvature might be a
``general property" of cosmological solutions of the classical Einstein 
equations.$^{28,29}$  This has aroused considerable controversy, more recently
acquiring the trendier name of gravitational chaos$^{29}$ and spreading into
higher dimensional purely gravitational models.$^{23-27,30,31}$
 
In the $d=4$ semisimple case, the vacuum supermomentum constraints restrict 
the spatial metric to be diagonalizable, the three spatial gauge degrees of
freedom being intimately connected with the remaining three ``offdiagonal 
modes" which are not excited in vacuum.  (Diagonalization depends on the 
choice of the Lie algebra basis; the Killing metric must be diagonal in a
basis in which the spatial metric is diagonal. Such a frame is then also an
eigenbasis of the extrinsic curvature, namely a ``Kasner frame"$^{28,46}$.)
In Taub time gauge the spatial curvature potentials at fixed $\alpha$ have
``essentially closed" potential contours on the flat subspace of the two
$\beta$ variables, and for increasing values of the potential, these contour
lines resemble more and more closely straight line contours joined together
at vertices which are either closed or have open channels which run out to
infinity with a width which goes to zero.  The ``straight" contours are
exponential and move outward with $\alpha$-speed $\half$ as $\alpha\to0$, 
while the system point moves with unit $\alpha$-speed (a null geodesic)
when the potential is negligible compared to the kinetic energy.  Thus the
system point continually overtakes and scatters off the receding exponential
potentials as long as the relative velocity between the system point in its
free state and some exponential potential is always negative (approach). 
In fact the
geometry is such that the relative velocity between the system point and the
exponential potentials which is is chasing
is nonpositive in every direction
but the vertex directions, where it vanishes.  Motion exactly along a vertex
direction only occurs for the Taub solutions which describe motion down the
center of an open channel extending to infinity.  Thus in general one has the
Mixmaster behavior$^{41}$ in which the system point rattles around 
indefinitely within the expanding potential as $\alpha\to0$.  On the other 
hand the nonsemisimple vacuum models have an open set of directions in which 
the system point may escape, i.e., the absence of at least one of the 
``straight" potential terms leads to an open set of directions in which
the potential asymptotically goes to zero, and the system point eventually
ends up in an asymptotic free state.
 
In the group manifold case, which is more general than the coset space 
case, each
``straight" potential contour line is associated with the structure constant
component $\C abc$ with $a\ne b\ne c$, which enters as a quadratic factor in
the potential.$^1$ For the $d=4$ semisimple case, all such terms are 
necessarily present, leading to the essentially closed potential contours.
In higher dimensions several crucial differences occur.  One still needs
semisimplicity for the existence of essentially closed potential contours;
these potentials are best described in a frame in which the Killing metric is
diagonal as in the $d=4$ case.  However, the Jacobi identity no longer allows
all structure constants $\C abc$ with distinct indices to be nonzero, removing 
certain potential terms.
 
Furthermore, only some of the offdiagonal modes are restricted by the 
supermomentum constraints, so the vacuum case in no longer diagonalizable and 
certain offdiagonal modes are nontrivial in the sense that the curvature
potential necessarily depends on them and they cannot be suppressed without
additional symmetry imposition, much like that which reduces the Mixmaster 
models to the essentially different Taub models.  This leads to the excitation 
of nonzero curvature modes of the (rescaled) DeWitt metric.  For those models
which are diagonalizable and therefore no longer general, the absence of all 
possible structure constants with distinct indices apparently
leads to a vertex geometry 
such that an open set of vertex directions exists along which the ``free"
system point can chase the potential with positive relative velocity 
(recession); the 
system point then eventually ends up in an asymptotic free 
state.$^{26,27}$
 
Without imposing the Jacobi identity, Demaret et al$^{30}$ have shown that 
this happens only for $d>10$, relevant to the general inhomogeneous case which
still remains somewhat controversial.  For the homogeneous case, the Jacobi
identities seem to 
reduce the critical dimension to $d>4$ in diagonalizable models.
However, in nondiagonal models, the eigenvectors of the extrinsic curvature 
no longer coincide with the time independent invariant frame vectors and the 
transformation to the eigenvector frame induces more nonzero structure 
functions.  Since it is the eigenvector frame (Kasner frame) structure 
functions which are relevant to the existence of an asymptotic free state, the
question still remains open for those dimensions $d\in[5,10]$ for which 
semisimple groups exist.
 
Of course chaos even in $d=4$ dimensions is sabotaged by the presence of a 
scalar field since it contributes a term to the super-Hamiltonian which 
reduces the free motion speed of the system point (timelike motion), 
therefore allowing an open 
set of vertex directions (or eventually all directions) which permit an 
asymptotic free state.$^{9,37}$   Moreover, since the free motion phases are
translations in $\alpha\sim \ln g$, one very quickly approaches the initial 
singularity $g=0$, slamming into the Planck scale where the classical Einstein 
equations break down and the indefinite approach to the classical singularity 
is no longer particularly relevant.  The classical chaos does lead to
quantum consequences, however, at least at the naive level which allows
calculation.$^{61}$
 
\section{Conclusion}
 
Whether one is interested in mathematical questions or more physical 
questions, a broader point of view is extremely valuable.  This article has 
focused on certain mathematical questions as a means of illustrating this 
point.  Staring at equations or their solutions in a 2-dimensional 
representation is not particularly illuminating.  The rich geometrical 
structure of the Einstein equations allows one instead to exploit ``hidden"
relationships and visualize many concepts which remain somewhat obscure in 
their brute form representation.  Even when visualization may be difficult or
not particularly relevant, a perspective which views a given problem ``from 
above" is certainly more effective in not only resolving that problem, but in 
understanding the result and placing it into context.  
The latter is, after all, more important than the 
former.

\section*{Acknowledgement}

The Vatican Observatory at Castel Gandolfo is thanked for its warm hospitality
during the period in which part of this work was done.  The I.N.F.N.
at the 
Department of Physics of the University of Rome
and the Exosat Observatory of Darmstadt, West Germany
are thanked for computer facilities used in
typesetting this manuscript with \TeX.
 
\section*{References}

\begin{enumerate}
\item R. T. Jantzen, \pr D{\bf34}, 424 (1986).
\item Th. Kaluza, {\it Sitz. Preuss. Akad. Wiss. Phys. Math.} K1,
966 (1921); O. Klein, {\it Z. Phys.} {\bf37}, 895 (1926).
\item B. DeWitt, {\it Dynamical Theories of Groups and Fields}, (Gordon and
Breach, New York, 1965), p.139.
\item R. Kerner, {\it Ann. Inst. H. Poincar\'e} {\bf9}, 143 (1968).
\item Y.M. Cho, \jmp {\bf16}, 2029 (1975).
\item Y.M. Cho and P.G.O. Freund, \pr D{\bf12}, 1171 (1975).
\item A. Salam and J. Strathdee, \ap {\bf141}, 316 (1982).
\item A. Chodos and S. Detweiler, \pr D{\bf21}, 2167 (1980).
\item V.A. Belinsky and I.M. Khalatnikov, \spj {\bf36}, 591 (1973).
\item P.G.O. Freund, \nucp B{\bf209}, 146 (1982).
\item P.G.O. Freund and M.A. Rubin, \pl {\bf97}B, 233 (1980).
\item F. Englert, \pl {\bf119}B, 339 (1982).
\item M.J. Duff, ``Supergravity, the seven-sphere and spontaneous symmetry
breaking" in {\it Proceedings of the Third Marcel Grossmann Meeting on
General Relativity} (North Holland, Amsterdam, 1983), p.269.
\item E. Cremmer and J. Scherk, \nucp B{\bf108}, 409 (1976),
B{\bf118}, 61 (1977).
\item E. Cremmer, B. Julia and J. Scherk, \pl {\bf76}B, 409 (1978).
\item J. Scherk and J.H. Schwarz, \pl {\bf82}B, 60 (1979).
\item J.F. Luciani, \nucp B{\bf135}, 111 (1978).
\item L. Castellani, L.J. Romans and N.D. Warner, \ap {\bf 157}, 394 (1984).
\item L. Castellani, \pl {\bf149}B, 103 (1984).
\item J. Demaret, J.-L. Hanquin, M. Henneaux and P. Spindel, \nucp {\bf252},
538 (1985).
\item J. Demaret and J.-L. Hanquin, \pr D{\bf31}, 258 (1985).
\item D. Lorentz-Petzold, \pl {\bf148}B, 43 (1984), {\bf149}B, 79 (1984),
{\bf151}B, 105 (1985), {\bf158}B, 110 (1985), {\bf167}B, 157 (1986);
\ptp {\bf73}, 533 (1985); \cqg {\bf2}, 829 (1985).
\item J.D. Barrow and J. Stein-Schabes, \pr D{\bf32}, 1595 (1985);
\pl {\bf167}B, 173 (1986).
\item P. Halpern, \pr D{\bf33}, 354 (1985).
\item T. Furusawa and A. Hosoya, \ptp {\bf73}, 467 (1985).
\item H. Ishihara, \ptp {\bf74}, 490 (1985).
\item A. Tomimatsu and H. Ishihara, \grg {\bf18}, 161 (1986).
\item V.A. Belinsky, I.M. Khalatnikov and E.M. Lifshitz, {\it Adv. Phys.}
{\bf31}, 639 (1985), {\bf19}, 225 (1970).
\item J. Barrow, {\it Phys. Rep.} {\bf85}, 1 (1982); D. Chernoff and J. 
Barrow, \prl {\bf50}, 134 (1983).
\item J. Demaret, M. Henneaux and P. Spindel, \pl{\bf164}B, 27 (1985).
\item J. Demaret, M. Henneaux, P. Spindel and A. Taormina, ``Fate of the
Mixmaster Behavior in Vacuum Inhomogeneous Kaluza-Klein Cosmological Models,"
preprint (1986).
\item A.H. Taub, {\it Ann. Math.} {\bf53}, 472 (1951).
\item C.W. Misner, K.S. Thorne and J.A. Wheeler, {\it Gravitation}, (Freeman,
San Francisco, 1973).
\item L. Bianchi, ``Sugli spazi a tre dimensioni che ammettono un gruppo
continuo di movimenti" (1897), in {\it Opere}, vol. {\bf9}, (Edizione 
Cremonese, Rome, 1958); {\it Lezioni sulla teoria dei gruppi continui di
trasformazioni} (1918), p.550.
\item R. Kantowski and R.K. Sachs, \jmp {\bf7}, 443 (1968).
\item F.B. Estabrook, H.D. Wahlquist and C.G. Behr, \jmp {\bf9}, 497 (1968).
\item R.T. Jantzen, \cmp {\bf64}, 211 (1978); \ncim {\bf55}B, 161 (1980); \jmp 
{\bf23}, 1137 (1982); \ap {\bf145}, 378 (1983); \pr D{\bf33}, 2121 (1986).
\item R.T. Jantzen, ``Spatially Homogeneous Dynamics: A Unified Picture," in
{\it Cosmology of the Early Universe}, R. Ruffini, Fang L.Z., eds. (World 
Scientific, Singapore, 1984) and {\it Gamov Cosmology}, R. Ruffini,
F. Melchiorri, eds. (North Holland, Amsterdam, 1987). 
\item C.W. Misner, ``Classical and Quantum Dynamics of a Closed Universe," in
{\it Relativity}, M. Carmeli, S.I. Fickler, L. Witten, eds. (Plenum Press, New 
York, 1970); ``Minisuperspace," in {\it Magic Without Magic}, J.R. Klauder, ed. 
(Freeman, San Francisco, 1972).
\item B.S. DeWitt, \pr {\bf160}, 1113 (1967).
\item  C.W. Misner, \prl {\bf22}, 1074 (1969); \pr {\bf186}, 1319 (1969).
\item R.T. Jantzen, \ap {\bf127}, 302 (1980).
\item D. Brill, \pr {\bf133}B, 845 (1964).
\item C.W. Misner and A.H. Taub, \spj {\bf28}, 122 (1969).
\item J.M. Bradley and E. Sviestens, \grg {\bf16}, 1119 (1984); E. Sviestens,
\grg {\bf17}, 521 (1985).
\item R.T. Jantzen, \jmp {\bf29}, 2748 (1986).
\item D. Brill and F. Flaherty, {\it Ann. Inst. H. Poincar\'e} A{\bf28}, 335 
(1978).
\item V. Joseph, {\it Proc. Camb. Phi. Soc.} {\bf62}, 87 (1966).
\item S.T.C. Siklos, \cmp {\bf58}, 255 (1978).
\item G.F.R. Ellis and M.A.H. MacCallum, \cmp {\bf12}, 108 (1969).
\item S.W. Hawking and G.F.R. Ellis, {\it The Large Scale Structure of 
Space-time} (Cambridge University Press, Cambridge, 1973).
\item D. Lorentz-Petzold, ``Exact Brans-Dicke-Bianchi Solutions," in {\it 
Lecture Notes in Physics} {\bf205}, 403, C. Hoenselaers, W. Dietz, eds. 
(Springer, Berlin, 1984).
\item K. Jacobs, \apj {\bf153}, 661 (1968).
\item D. Kramer, H. Stephani, M.A.H. MacCallum and E. Herlt, {\it Exact 
Solutions of Einstein's Field Equations} (VEB, Deutscher Verlag der 
Wissenschaft, Berlin, 1980).
\item O. Bogoyavlensky and S.P. Novikov, \spj {\bf37}, 747 (1973);
{\it Russian Math. Surveys} {\bf31}, 31 (1976); {\it Proc. I.G. Petrovsky 
Seminar}, 7 (1973): English translation in {\it Sel. Math. Sov.} {\bf2}, 159 
(1982).
\item O. Bogoyavlensky, \spj {\bf43}, 187 (1976).
\item A.A. Peresetsky, {\it Russian Math. Notes} {\bf21}, 39 (1977).
\item K. Rosquist, \cqg {\bf1}, 81 (1983).
\item J. Wainwright, \grg {\bf16}, 657 (1984).
\item R.T. Jantzen and K. Rosquist, \cqg {\bf3}, 281 (1986).
\item T. Furusawa, ``Classical and Quantum Chaos of the Mixmaster Universe" 
and ``Quantum Chaos of the Mixmaster Universe," preprints (1986).

\end{enumerate}

Later work in this direction:

\begin{itemize}

\item 
R.T. Jantzen, 
\PR \vol{D35}, 2034 (1987)\\{} [errata for Reference 1 and addendum].

\item
R.T. Jantzen, 
\PL \vol{186B}, 290 (1987).

\item
R.T. Jantzen,
\PR \vol{D37}, 3472 (1988).

\item
C. Uggla, K. Rosquist and R.T. Jantzen, 
\PR \vol{D42}, 404 (1990).

\item
R.T. Jantzen and C. Uggla, 
\GRG \vol{24}, 59 (1992).

\item
C. Uggla, R.T. Jantzen and K. Rosquist, 
\GRG \vol{25}, 409 (1993).   
 
\end{itemize}

Abstracts available at:
 
{\tt http://www.homepage.villanova.edu/robert.jantzen }

\end{document}